\documentclass{article}

\usepackage{PRIMEarxiv}

\usepackage[utf8]{inputenc} 
\usepackage[T1]{fontenc}    
\usepackage{hyperref}       
\usepackage{url}            
\usepackage{booktabs}       
\usepackage{amsfonts}       
\usepackage{nicefrac}       
\usepackage{microtype}      
\usepackage{lipsum}
\usepackage{fancyhdr}       
\usepackage{graphicx}       
\graphicspath{{media/}}     
\usepackage{amsmath}
\usepackage{caption}
\usepackage{subcaption}
\usepackage{tabularx}
\usepackage{numprint}
\usepackage{graphicx}
\usepackage{xcolor}
\usepackage{textcomp}
\usepackage{capt-of}
\usepackage[linesnumbered,vlined,ruled]{algorithm2e}
\usepackage{cleveref}
\npthousandsep{,}\npthousandthpartsep{}\npdecimalsign{.}

\def\HiLi{\leavevmode\rlap{\hbox to \hsize{\color{yellow!50}\leaders\hrule height .8\baselineskip depth .5ex\hfill}}}
\newcommand{\pluseq}{\mathrel{+}=}

\title{Evaluating the Feasibility of a Provably Secure Privacy-Preserving Entity Resolution Adaptation of PPJoin using Homomorphic Encryption
}

\author{
  Tanmay Ghai, Yixiang Yao, Srivatsan Ravi, and Pedro Szekely \\
  \texttt{\{tghai\}@isi.edu, \{yixiangy, srivatsr, szekely\}@usc.edu} 
}

\begin{document}
\maketitle

\begin{abstract}
Entity resolution is the task of disambiguating records that refer to the same entity in the real world. In this work, we explore adapting one of the most efficient and accurate Jaccard-based entity resolution algorithms - PPJoin, to the private domain via homomorphic encryption.
Towards this, we present our precise adaptation of PPJoin (HE-PPJoin) that details certain subtle data structure modifications and algorithmic additions needed for correctness and privacy. We implement HE-PPJoin by extending the PALISADE homomorphic encryption library and evaluate over it for accuracy and incurred overhead. Furthermore, we directly compare HE-PPJoin against P4Join, an existing privacy-preserving variant of PPJoin which uses fingerprinting for raw content obfuscation, by demonstrating a rigorous analysis of the efficiency, accuracy, and privacy properties achieved by our adaptation as well as a characterization of those same attributes in P4Join.
\end{abstract}


\section{Introduction}
\label{sec:intro}

\begin{figure}[!hbt]
    \centering
    \includegraphics[width=.6\linewidth]{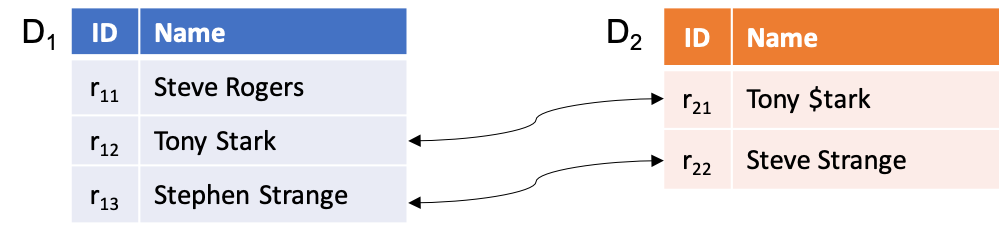}
    \caption{An example of entity resolution. Pairs ($r_{12}, r_{21}$) and ($r_{13}, r_{22}$) refer to the same entities in the real world.}
    \label{fig:er}
\end{figure}

The \textit{entity resolution} problem involves finding pairs across datasets that belong to different owners which refer to the same entity in the real world. For example, consider two datasets (e.g. $D_1, D_2$) with distinct collections of records (e.g. \{$r_{11}, r_{12}, r_{13}$\}, \{$r_{21}, r_{22}$\}) containing representative information about \textit{The Avengers} as detailed in \cref{fig:er}. The task here is to determine which pairs between $D_1$ and $D_2$ correspond to the same superheroes. As shown in the figure, ($r_{12}, r_{21}$) both refer to ``Tony Stark'' (commonly known as Iron Man), while ($r_{13}, r_{22}$) correspond to ``Stephen Strange'' (or Dr. Strange). There exist many traditional entity resolution algorithms that, in many cases, directly compare each pair of records among datasets resulting in quadratic time complexity. Among such approaches, the PPJoin (Position Prefix Join)~\cite{xiao2011efficient} algorithm is one that seeks to be more \textit{efficient} by pruning the record comparisons as part of the search space via filtering procedures. In addition to efficiency, one common concern for applications of entity resolution where the datasets may contain sensitive information (e.g. medicine, financial institutions) is that of \textit{privacy}. In such cases, the goal is to perform entity resolution while not revealing any identifiable or sensitive data content to any data owners or to an adversary; the problem to solve entity resolution given this additional setting is known as \textit{privacy-preserving entity resolution}. P4Join~\cite{sehili2015privacy} is an existing privacy-preserving variant of PPJoin established to provide privacy utilizing encrypted bit vectors. While P4Join attempts to hide the raw content of data it operates over, it remains susceptible to possible attacks and mis-configurations (discussed in \cref{sec:p4join-security}) that can reveal sensitive information to adversaries. Thus, we contend that the security it provides is \emph{not} sufficient in real-world settings.

We therefore investigate a more \emph{provably} secure adaptation of PPJoin into the privacy-preserving setting using homomorphic encryption~\cite{cryptoeprint:2019:939}. Towards this, we implement new data structure operations and perform algorithmic modifications to establish our adapted version: HE-PPJoin; we further evaluate over real-world data and settings given an adversary model, and rigorously compare our adaptation against P4Join for efficiency, accuracy, and security.

\noindent\textbf{Summary of our contributions}
In this paper, we propose a homomorphic adaptation of the PPJoin algorithm, along with full pseudocode detailing explicitly the necessary algorithmic changes, as well as an empirical analysis profiling the run time performance and accuracy against real world data. To do this, we \textit{invent} an $n$-party protocol analagous to PPJoin, that maintains its correctness and efficiency properties, while also guaranteeing \textit{no} leakage of fatally sensitive information (see \cref{sec:realworldprivacy} for a precise discussion of privacy leakage). While our algorithmic changes obfuscate the underlying record content being operated over, we ensure that there is additionally \textit{no} accuracy loss by implementing private operators that are deterministic in their output. Finally, we develop a P4Join Python implementation and provide a rigorous comparison against it, detailing possible security flaws, running time vs. accuracy trade-offs, and volatility present in its approach.  

\noindent\textbf{Structure of the paper}: \Cref{sec:problemdef} formalizes both the entity resolution and privacy-preserving entity resolution problems, \cref{sec:preliminaries} details the motivation for our work, as well as an in-depth introduction into PPJoin, P4Join, and homomorphic encryption, \cref{sec:he-adaptation} formally presents our adaptation, its pseudocode, as well proofs of correctness \& privacy, \cref{sec:experiments} reports the settings and findings of our experiments (and comparisons against P4Join), and finally \cref{sec:conclusion} summarizes and presents directions for future work.

\section{Problem definition}
\label{sec:problemdef}

The \textbf{entity resolution} (ER) problem can be described as an algorithmic set $P=(D_1 \ldots D_n,\allowbreak M)$ where $D_1 \ldots D_n$ are different datasets consisting of a collection of unique records, from different data owners or parties $P_1 \ldots P_n$. $M$ denotes the matching record pairs ($r_i, r_j$ denoting record ids) between any two datasets amongst the $n$ parties, that is, $M=\{(r_i, r_j)\,|\,r_i = r_j; r_i \in D_k, r_j \in D_l\}$, where $r_i=r_j$ indicates that $r_i$ and $r_j$ refer to the same entity in the real world, and $D_k, D_l$ are any two datasets $\in$ \{$D_1 \ldots D_n$\}. 

Normally, $n \in [1, \infty)$. When $n=1$, the ER task is within one dataset. This particular problem is called \textbf{de-duplication}. The $M$ in this case is, $M=\{(r_i, r_j)\,|\,r_i = r_j; r_i, r_j \in D\}$, where both $r_i$ and $r_j$ come from the same dataset $D$. If $n \geq 2$, we focus only on resolving entities across datasets. While we define ER here to be applicable in the $n$ party setting, \cref{sec:intro} introduces it while focusing on the 2-party case. We intentionally make this decision to motivate privacy-preserving ER with a real-world example. 

The \textbf{privacy-preserving entity resolution} (PPER) problem extends ER by requiring the original data to be encoded or masked in such a way that it is not exposed to entities part of the computation model or to an adversary. In general, there are many methods that can be used to do this (i.e. hashing, bit-vector encodings, embeddings, etc), but for our purposes we adopt encryption as our tool to obfuscate. Keeping the same structure as with ER, here, $P_1 \ldots P_n$ encrypt their records and send $Enc(D_1) \ldots Enc(D_n)$ to a computation host to determine $M'$, which is analogous to $M$, but over encrypted data: $M'=\{(r_i, r_j)\,|\,r_i = r_j, r_i \in Enc(D_k), r_j \in Enc(D_l)\}$.
An additional requirement here is that the probability $P_1 \ldots P_n$ learn any information outside of their dataset ($D_i$) or the matching ($M'$) is negligible, and the probability that a computation host learns anything outside of $M'$, including anything in or about \{$D_1 \ldots D_n$\}, is negligible amongst the presence of adversaries.

\section{Motivation \& Preliminaries}
\label{sec:preliminaries}
This section establishes the motivation behind our work, provides a detailed walk-through of PPJoin~\cite{xiao2011efficient} and P4Join~\cite{sehili2015privacy} contextualized with the real-world example from \cref{sec:intro}, and introduces our main privacy tool: homomorphic encryption~\cite{cryptoeprint:2019:939}, while also providing a rigorous security analysis of P4Join.

\subsection{Motivation}
To determine whether two records $(r_i, r_j)$ form a pair in the real world, algorithmically, string similarity metrics can be used to score how similar they are based on their representative information. These records form a \textit{candidate pair}, which is considered to be a \textit{true} pair iff. the final similarity score is above a certain threshold.

Among various string similarity metrics, \textbf{Jaccard similarity}~\cite{jaccard1901etude} is well-known and widely used.
Given two sets of tokens $x$ and $y$, the Jaccard similarity between them can be described as the size of the intersection divided by the size of the union, that is, $Sim_{Jaccard}(x, y) = \frac{|x \cap y|}{|x \cup y|} = \frac{|x \cap y|}{|x| + |y| - |x \cap y|}$. Since the inputs of Jaccard similarity are sets, the raw strings as part of records need to be tokenized (e.g. \textit{n}-grams~\cite{banerjee2003design}). Therefore, $(r_i, r_j)$ forms a pair when $Sim_{Jaccard}(x, y) \geq t$ where $x=tokenize(r_i)$, $y=tokenize(r_j)$ and $t$ is the pre-set threshold for minimum similarity. For the datasets that $r_i, r_j$ are a part of (e.g. $D_k, D_l$), such comparison ought to be over \textit{all} possible candidate pairs. We define this notion to be a  \textbf{full comparison} between $D_k$ and $D_l$: $T=\{(r_i, r_j)\,|\,r_i\in D_k, r_j \in D_l\}$, and the total number of candidate pairs to be $|T|$, the Cartesian product between sizes of $D_k$ and $D_l$, i.e. $|D_k| \times |D_l|$. 

As an example, consider $r_{12}, r_{21}$ from \cref{fig:er} tokenized into \textit{bi}-grams during PPJoin pre-processing (\cref{fig:ppjoin_preprocessing}). The Jaccard similarity between the two sets, \{\texttt{on}, \texttt{ar}, \texttt{ny}, \texttt{y-}, \texttt{rk}, \texttt{ta}, \texttt{st}, \texttt{to}, \texttt{-s}\} and \{\texttt{ar}, \texttt{on}, \texttt{to}, \texttt{ny}, \texttt{y-}, \texttt{rk}, \texttt{ta}, \texttt{\$t}, \texttt{-\$}\}, is $7/11$ since the size of their intersection is 7, while the size of their union is 11. 

Looking for a needle (a true pair) in a haystack (set of all candidate pairs) is not easy, and usually, this process is computationally heavy; exhausting quadratic full comparison for finding true pairs is \emph{not} practical in the real world  for it is \emph{not} scalable. For such reasons, pruning $T$ by removing pairs that are unlikely to be part of $M$, \emph{before} comparison, with relatively low-budget methods is far more feasible. Blocking~\cite{michelson2006, 10.1145/3459637.3482318} is one category of such methods. However, the benefit of blocking also comes with an inherent side-effect: true pairs might not be thoroughly recognized, and thus discarded. This potential problem requires a non-trivial amount of tuning in practice to be effective. Unlike blocking, PPJoin, is an algorithm specifically designed to exploit features of Jaccard similarity in order to avoid a full comparison. Though it's complexity is still quadratic in the worst case, in \emph{many} instances it is able to successfully and efficiently filter out candidate pairs that are not likely to be in $M$ \textit{without} sacrificing true pairs part of $M$~\cite{10.14778/2732296.2732299, 10.14778/1920841.1920904}. We walk-through PPJoin in detail in \cref{sec:ppjoin}. 

P4Join is a privacy-preserving variant of PPJoin; it adapts PPJoin's filtering to be applied over encoded bit vectors by hashing the original records into signatures. This prevents parties \emph{other than} the data owner from seeing records in clear and obfuscates the raw content of records, however, it only achieves preliminary security, susceptible to attacks where an adversary can recover sensitive record content. In particular, attacks such as \textit{enumeration}, bit-sensitivity \textit{simulation}, as well as parameter \textit{misconfiguration} and \textit{rainbow table} analysis are all in play given P4Join's usage of Bloom-filters for one-way hashing. Furthermore, a level of accuracy loss is incurred due to a fixed-length requirement of records in P4Join. In order to overcome such issues, we \textbf{motivate} our idea to adapt PPJoin utilizing homomorphic encryption for better security guarantees while achieving full accuracy; \cref{sec:he} describes our scheme of choice and it's underlying ideas and properties. A walkthrough of P4Join is present in \cref{sec:p4join}, and an analysis of its security properties and susceptibility to possible attacks in \cref{sec:p4join-security}.

\subsection{PPJoin}
\label{sec:ppjoin}

\begin{figure*}[!t]
    \includegraphics[width=0.9\linewidth]{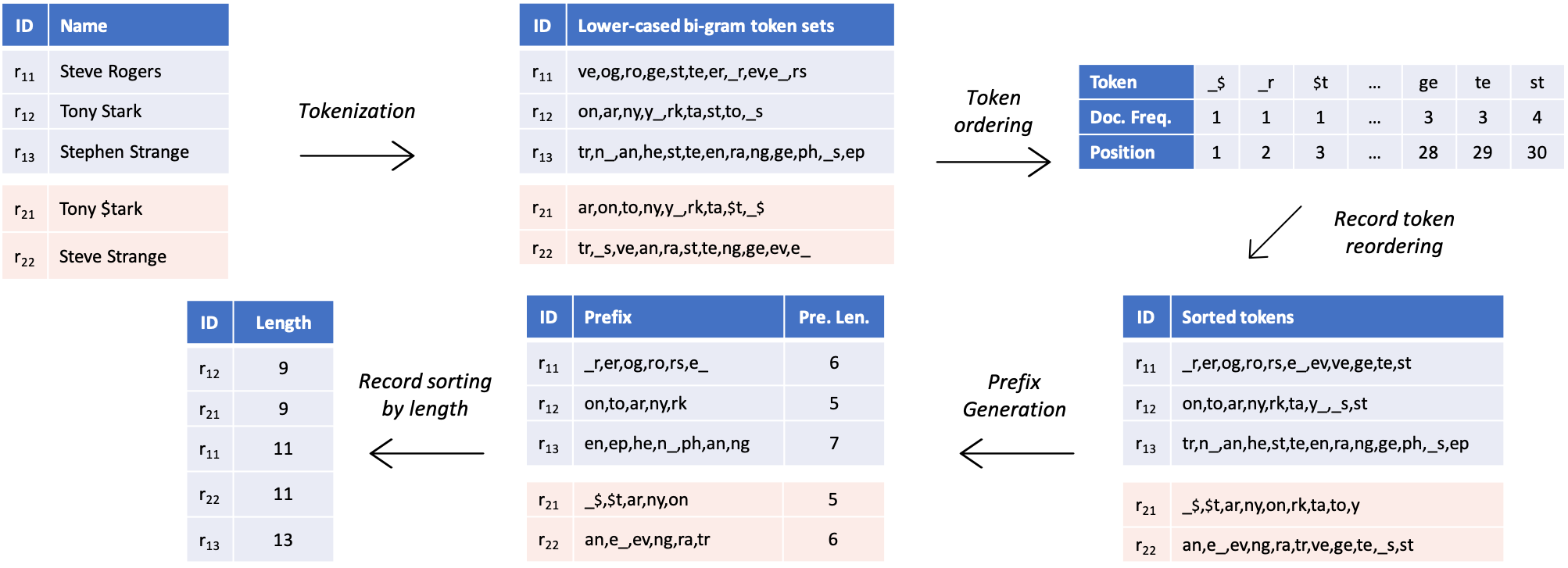}
    \centering
    \caption{PPJoin pre-processing. Record tokenization (to bi-grams) begins the process, followed by token counting and positional ordering in a document frequency map. This map is then used to reorder tokens within records according to their frequency-based position. Then, prefixes for each record are generated, and all records are sorted in a multi-set based on ascending length. }
    \label{fig:ppjoin_preprocessing}
\end{figure*}

PPJoin (Position Prefix Join) \cite{xiao2011efficient} is an extended implementation of Jaccard similarity that addresses the Cartesian product shortcoming of the naive implementation by drastically reducing the number of candidate pairs that must be considered, leading to much improved efficiency. The intuition behind PPJoin is to avoid comparison between record pairs that have huge length differences or an overlap of tokens that is below some pre-defined threshold. 

\noindent\textbf{Definitions}: The filtering methods of PPJoin are derived from the basic Jaccard definition. Before detailing them, two additional definitions are needed first: $Overlap(x, y) = |x \cap y|$ and $Hamming(x,y) = |(x - y) \cap (y - x)|$.
$Overlap(x, y)$ is the size of the intersection between two records or token sets ($x$ and $y$); it is additionally the numerator in Jaccard similarity. $Hamming(x,y)$ is the minimum number of substitutions required to change one string into another. The constraint defined with the presence of a threshold $t$ can then be transformed into the following equivalent forms: $Sim_{Jaccard}(x, y) \geq t \Leftrightarrow Overlap(x, y) \geq \lceil \frac{t}{1+t} \cdot (|x| + |y|) \rceil = \alpha$, $Overlap(x,y) \geq \alpha \Leftrightarrow Hamming(x,y) \leq |x| + |y| - 2\alpha$, and $Sim_{Jaccard}(x, y) \geq t \Rightarrow t \cdot |x| \leq |y|$. Among them, the first one can be seen as \textit{length filtering}. Moreover, according to the third one, pairs of records with an overlap less than the minimal overlap $\alpha$ cannot meet the similarity join condition. In addition to length filtering, PPJoin employs \textit{prefix filtering}. The intuition here is that similar records have some fragments that overlap with each other~\cite{chaudhuri2006primitive}. The prefix $pref(x)$ of a record $x$ is computed as $|pref(x)| = \lceil (1-t) \cdot |x| \rceil + 1$. In conjunction with the Hamming overlap inequality, it follows that $(|x|-\alpha+1)-pref(x)$ of $x$ and the $(|y|-\alpha+1)-pref(y)$ of $y$ must share at least one token. Lastly, PPJoin utilizes \textit{position filtering}. The key idea here is that a particular token $\omega$ splits records $x$ and $y$ ($\omega \in x \cap y$) into left ($x_l(\omega)$ for $x$ and $y_l(\omega)$ for y) and right partitions ($x_r(\omega)$ for $x$ and $y_r(\omega)$ for y), where tokens in each record are sorted by a particular ordering $O$. It then follows that $Overlap(x,y) \geq \alpha \Rightarrow Overlap(x_l(\omega),y_l(\omega))+min(|x_r(\omega)|,|x_r(\omega)|) \geq \alpha$. This forms the basis for determining if two records are similar based on common tokens in their prefixes and \textit{unseen} right partitions. 

\noindent\textbf{Pre-processing}: Before applying the filtering procedures, data must be pre-processed; the steps for this are decomposed in \cref{fig:ppjoin_preprocessing}. Records are firstly tokenized into token sets using data dependent methods: here we utilize lower-case, \textit{bi}-gram tokenization. Additionally, in order to determine if the overlap between two records is adequate when executing PPJoin, tokens and records need to be processed in a certain order. Therefore, tokens from each record are counted (document frequency) and sorted in ascending order (assigning each a position) in a global token ordering $O$. If the frequency of two tokens are the same, they (amongst them) can be positioned arbitrarily as long as the relative order is consistent throughout the computation. Utilizing the newly assigned positions in $O$, tokens in each record are re-ordered accordingly. The last two steps are in preparation for the filtering procedures: for prefix filtering, each record's prefix length is calculated according to $pref(x)$; for length filtering, the records are sorted by their lengths.

\noindent\textbf{Algorithm}: The PPJoin algorithm is described in two parts: algorithm 1 (\texttt{ppjoin}) and 2 (\texttt{Verify}) in \cite{xiao2011efficient}. It has three phases: \textit{indexing} for indexing tokens into an inverted index, \textit{candidate generation} for generating candidate pairs probed on tokens in the prefixes of records from the inverted index, and \textit{verification} for verifying if a candidate pair satisfies the overlap constraint.

In \texttt{ppjoin}, PPJoin constructs the inverted index by inserting tokens from records utilizing their positions. Additionally, a hash map is generated for accumulating overlaps between records. For each record, its prefix is compared against the prefix of another to check if they both contain the same token, and the length filter is applied over them. The satisfied record pair is tested by the prefix and position filters. The overlap stored in hash map is incremented if the current overlap plus the upper bound (maximum possible overlap of the right partition) is greater than equal to $\alpha$.

For each record, the hash map contains the number of overlapping tokens between all other records. \texttt{Verify} then verifies if the candidate pair have enough overlap to be considered as true pair. The heuristic is to use the last token in the prefix of each record as a pivot, where only the record with the smaller token in the suffix needs to be intersected with the other record. If the resulting intersection count plus the original overlap is still greater than $\alpha$, the pair is added to the set of true pairs.

\subsection{Privacy-preserving PPJoin: P4Join}
\label{sec:p4join}

P4Join (Privacy-Preserving Prefix Position Join)~\cite{sehili2015privacy} is a privacy-preserving variant of PPJoin, converting records into fingerprints represented as same-sized bit arrays.

\noindent\textbf{Definitions} The basic hashing method $h_i$ for record encryption is defined as: $h_{i}(x) = (f(x) + i \cdot g(x)) \texttt{ mod } l$, where $x$ is a set of a $n$-gram tokens generated from a record, $f$ and $g$ are primitive hash functions and $l$ is the length of the fingerprint. Specifically, $f$ is HMAC-SHA1 and $g$ is HMAC-MD5. Each function $h_{i}$, where $i$ is from 1 to $k$, sets one bit in the bit array. Each token is hashed into a bit array by applying $h_i$ multiple times, thus a token is represented by the $k$ set bits, all corresponding to the final fingerprint.

As an example, applying $bi$-gram tokenization to the record \texttt{Tony} will produce: \texttt{\{\_t, to, on, ny, y\_\}}. If $k=2$ and $l=10$, the token \texttt{\_t} can be mapped to bits 3 and 5 by applying $h_{1}$ and $h_{2}$ respectively, resulting as \texttt{[0,0,1,0,1,0,0,0,0,0]}. This method is applied on all following tokens to complete computing the bit array. Given this new representation, Jaccard similarity can still be applied with slight variations; known as Tanimoto similarity~\cite{tanimoto1958elementary}: $Sim_{Tanimoto}(x, y) =  \frac{|x \wedge y|}{|x \vee y|} = \frac{|x \wedge y|}{|x| + |y| - |x \wedge y|}$, where $|x|$ and $|y|$ denotes the number of set bits (cardinality) in the bit array $x$ and $y$ respectively.

\noindent\textbf{Pre-processing}: In very similar fashion as the pre-processing of PPJoin, P4Join's pre-processing also requires tokens to be reordered, prefixes to be calculated and records to be sorted according to length. With Tanimoto similarity, token frequency is calculated based on set bit positions; token reordering is then done according to frequency in ascending order. Prefixes here, additionally, refer to the number of set bits as well, as opposed to the size calculated in $pref(x)$. And lastly, records are sorted by ascending set-bit cardinality.

\noindent\textbf{Algorithm}: In its algorithmic procedure, P4Join does not construct an inverted index because the cost of maintenance outweighs the achievable savings on bit arrays. Instead, it maintains a mapping from set-bit cardinality to all relevant record ids. Then, the three filters are optionally applied on the candidate pairs; pairs that pass them are evaluated via Tanimoto similarity~\cite{tanimoto1958elementary}.

One incurred penalty introduced in P4Join is that of \textit{accuracy}: while PPJoin is a loss-less approach that produces exactly the same result as Jaccard similarity, P4Join is not. The reason behind this is P4Join leverages \textit{Bloom-filter} like encoding methods that one-way hash the records into \textit{fixed-length} fingerprints no matter the length of the originating record. Furthermore, tuning the parameters $k$ and $l$ with large amounts of data is a \textit{non-trivial} task. Though \cite{sehili2015privacy} does not suggest how to choose these parameters optimally, real world implementations using Bloom-filters~\cite{luo2018optimizing, medjedovic_2022} give us the following false positive rate: $f \approx (1-e^{-\frac{Uk}{l}})^k$, where $U$ indicates the total number of unique tokens. Thus, inducing the optimal $k$ to be $k_{opt} = \frac{l}{U}ln2$.
A lower false positive rate $f$ provides better linkage quality, but in turn provides poorer privacy. So, even if $k$ and $l$ can be adjusted to an optimal combination, hash collisions remain an \textit{unavoidable} issue due to the level of obfuscation needed for privacy-preservation. Moreover, the improved efficiency provided by the three filters in PPJoin is \textit{not} realized in P4Join, and in many instances applying said filters can be \textit{worse} than running a full comparison of Tanimoto similarity. Specifically, the overhead incurred by the position filter noticeably slows down execution. We perform thorough experiments regarding parameter tuning and filtering efficiency over our own implementation of P4Join; results for these are presented in \cref{sec:p4join-experiments}.
\subsubsection{Security Analysis and possible attacks}
\label{sec:p4join-security}
Though P4Join seeks to introduce privacy protections as they pertain to PPJoin, the effects are \textit{limited}, leaving record data vulnerable to possible attacks. 

In particular, P4Join employs one-way hashing via Bloom-filter encodings as its main privacy tool. While such a technique may be an efficient (or first-pass) attempt at privacy-protection, it does not provide any \textit{strong} privacy guarantees and is \textit{provably insecure} in certain scenarios. For example, in cases where the universal set of elements (in this case tokenized \textit{bi}-grams) is \textit{enumerable}, a simple \textbf{brute-force} attack is possible where an adversary with sufficient computational resources can simply check each possible element and reconstruct back the filter's content~\cite{bloomfilterprivacy}. In such a case, the false-positive rate ($f$) acts as the \textit{only} (weak) barrier for privacy protection, introducing ambiguity in the enumeration process.

Another form of attack that P4Join is susceptible to is that of \textbf{simulation}. Building off analysis in \cite{durham2013composite,vatsalan2016multi}, when dealing with bit-array fingerprints, the sensitivity of each bit can be learnt, directly revealing the underlying hashed content. Specifically, if we denote a bit position as $\beta_x$, we have the following: $dist(\beta_x) = |t|: \forall t \in U, h_i(t) = \beta_x, 1 \leq i \leq k$; $freq(\beta_x) = |r|: \forall r \in D, \beta_x\ in\ r = 1$; and $S(\beta_x) = 1 / min(dist(\beta_x), freq(\beta_x))$, where $1 \leq x \leq l$. In these equations, $dist(\beta_x)$ indicates the distribution of a certain token that appears in position $\beta_x$ and the $freq(\beta)$ is the number of records ($|r|$) in the dataset $D$ which have the bit set at $\beta_x$. The sensitivity of $\beta_x$, $S(\beta_x)$, is defined as the reciprocal of the minimal value between $dist(\beta_x)$ and $freq(\beta_x)$. Following this, if a bit position ($\beta_x$) represents a large number of tokens, and if these tokens correspond to the same record (i.e. higher sensitivity), the more vulnerable the content of the corresponding record will be~\cite{vatsalan2016multi}. In this way, an adversary can easily guess content present in the underlying records via an understanding of the sensitivity of certain bit positions.  

\textbf{Mis-configuration} of parameters is another natural issue that P4Join suffers from, following the two previously discussed above (specifically in terms of the false positive rate $f$). As mentioned, the parameters $k$ and $l$ need to be selected properly as their relationship is inversely proportional, representing the trade-off between accuracy and privacy as explained in calculating false positive rate $f$ and optimal $k$. For example: $k$ and $l$ are both integers, where $k << l$. In this scenario, it is highly possible (if not likely) that hashes for each record represent a unique fingerprint. Given uniqueness, even passive adversaries who just monitor data transmission channels can easily figure out which bits represent which tokens via simple cross-checking (e.g. eavesdropping). Furthermore, in active adversary (e.g. man-in-the-middle) settings this problem can be exacerbated if the computation unit is hacked or independent connections are formed for relaying impersonated messages. In the reverse scenario ($k$ is larger than $l$), the strength of privacy protection improves, however, the accuracy of the algorithm (as claimed) is in turn sacrificed. 

Finally, since the fingerprint generation process in P4Join is not aided with any random noise or salt, the same record content is guaranteed to generate the exact same fingerprint every time. With enough time for information collection, an adversary could easily map fingerprints back to clear-text via the construction of a \textbf{rainbow table} for low-cost future attacks. This in combination with the unprotected communication channels between parties can lead to easy data sniffing, interception, and even manipulation. 

\noindent \textbf{Remark}: Many of the security issues discussed are based on P4Join's use of Bloom-filters for one-way hashing record content into fixed-size signatures. In previous works, some cryptographic extensions (e.g. ~\cite{goh03,cryptobloom, pesusingbloom, 10.1145/1655008.1655025}) of Bloom filters have been suggested for stronger privacy requirements~\cite{bloomfilterprivacy}. While these options could have been explored, they remain limited in their practicality and implementation (library) support.

\subsection{Homomorphic encryption}
\label{sec:he}

In this section we provide the necessary background to discuss the privacy properties in our adaptation.

\subsubsection{The BGV scheme}
We work with the BGV~\cite{cryptoeprint:2011:277} (Brakersi, Gentry, Vaikuntanathan) fully homomorphic construction. BGV is an ideal choice for our use case due to its ability to support SIMD (single instruction multiple data) operations over its plaintexts and integer arithmetic over the learning with errors (LWE)~\cite{5497885} instance, as well as the corresponding ring (RLWE~\cite{cryptoeprint:2012:230}) instance. Homomorphic encryption~\cite{cryptoeprint:2019:939} can be summarized as a quartet of algorithms ($KeyGen$, $E$, $D$, $f$) where $KeyGen$ outputs a key-pair $(pk, sk)$ containing a public key ($pk$) and a private key ($sk$), $E$ is an encryption algorithm that takes as input the public key $pk$, message $m$, and outputs the ciphertext $c$, $D$ is a decryption algorithm that takes as input the private key $sk$, ciphertext $c$ and outputs the message $m$, and finally, $f$ is an evaluation function evaluated over ciphertexts outputting $c_f$ (an encrypted computation). The \textit{homomorphic} property of HE refers to the upholding of the following: $E(m_1) * E(m_2) = E(m_1 * m_2)$ $\forall m_1, m_2 \in M$, where $*$ represents a homomorphic operation, and $M$ represents the plaintext space~\cite{hetheoryapplication}. 

\noindent \textbf{Security properties}: HE schemes, are proven to be \textit{semantically}~\cite{Sako2011} and IND-CPA~\cite{646128, GOLDWASSER1984270} secure given the following scenario: provided two equally likely, distinct messages $m_1, m_2$ and a ciphertext $c$, no adversary should have an advantage in guessing (with probability $p > \frac{1}{2}$), whether $c$ is an encryption of $m_1$ or $m_2$~\cite{securityofhe}. Furthermore, with respect to \textit{semi-honest} adversaries~\cite{Hazay2010SemihonestA}, the BGV scheme can be extended to provided \textit{evaluation privacy}; this refers to the security and preservation of the operations applied over a ciphertext $c$, such that an adversary cannot tell what was performed to obtain $c$.  

\noindent \textbf{Remark}: Li and Micciancio in \cite{cryptoeprint:2020:1533} have recently shown that the IND-CPA security model is insufficient for approximation encryption schemes (e.g. CKKS~\cite{heforarithmeticofapproximatenumbers}), due to leakage of LWE noise in the approximation error potentially revealing the secret key. This is particularly harmful in scenarios where two or more parties encrypt their data under a jointly established public key and send it to a computation host to compute over the joint sets. Since BGV utilizes an exact decryption algorithm, the above notion of IND-CPA security still applies, and it is not susceptible to such attacks.

\subsubsection{Threshold HE}
While PPJoin isn't inherently a distributed protocol, translating it to be \emph{private} naturally requires secure multi-party computation~\cite{yao1982protocols}. Motivating the use of HE in such scenarios can be done with the help of threshold-HE~\cite{Desmedt2011, Schoenmakers2011}.

In threshold-HE, $KeyGen$, $E$, and $D$ in the quartet of algorithms used to describe HE are replaced by distributed key generation ($DKG$), distributed encryption ($DE$), and distributed decryption ($DD$) protocols, converting key generation and decryption into \textit{interactive} processes. In particular, $DKG$ generates a public $(pk)$, secret $(sk)$ key-pair, where $sk$ is split and distributed into $n$ secret shares, one-per-party. $DE$ uses $pk$ on behalf of all parties, and $DD$ is done collectively by merging the results of each party's individual decryption. The threshold parameter $d$ (where $d \leq n$) determines how many secret shares of $sk$ are needed to correctly perform $DD$. 

\noindent \textbf{Security properties}: Threshold homomorphic encryption is also provably \textit{semantically} secure, given the following scenario: provided $n$ parties and a secret key $sk$, no adversary, given secret key shares ${sk}_{i \in S}$ for a set $S$ with a dimension $ < d$ (where $d \leq n$), should gain any additional advantage in recovering $sk$~\cite{cryptoeprint:2017:257}.

\section{HE-PPJoin}
\label{sec:he-adaptation}

\begin{algorithm}[t]
\caption{Helper functions}
\label{alg:helper}
\SetKwInput{KwInput}{Input}
\SetKwFunction{KeyGen}{KeyGen}
\SetKwFunction{GenBGV}{GenBGV}
\SetKwFunction{MultipartyKeyGen}{MultipartyKeyGen}

\SetKwFunction{IsMatch}{IsMatch}
\SetKwFunction{Encrypt}{Encrypt}
\SetKwFunction{Decrypt}{Decrypt}

\SetKwFunction{DocFreqJoin}{DocFreqJoin}

\SetKwFunction{PrivateSetIntersection}{PrivateSetIntersection}

 \SetKwProg{Fn}{Function}{:}{}
 \KwIn{$kp_1 \ldots kp_n$: ($pk_i, sk_i$) key-pairs for $P_1 \ldots P_n$}
 \KwOut{$kp_{mp}$: multiparty key-pair}
    \Fn{\KeyGen{$kp_1 \ldots kp_n$}}{
        $kp_{mp} \gets \MultipartyKeyGen{$kp_1 \ldots kp_n$}$; \\
        \KwRet $kp_{mp}$;
    }
    \BlankLine
     \KwIn{$kp_{mp}$: multiparty key-pair; $Enc(t_i), Enc(t_j)$: encrypted tokens; $r_*$: randomly sampled non-zero ciphertext}
     \KwOut{boolean determining if $t_i, t_j$ are equal}
    \Fn{\IsMatch{$kp_{mp}$, $Enc(t_i), Enc(t_j)$}}{
        $sub \gets (Enc(t_i) - Enc(t_j)) \cdot r_*$; \\
        \KwRet $(\Decrypt{$kp_{mp}, sub$} == 0)$;
    }
    \BlankLine
    \KwIn{$kp_{mp}$: multiparty key-pair; $O_1 \ldots O_n$: local (encrypted) document frequency mappings}
    \KwOut{$O$: coalesced global (encrypted) mapping}
    \Fn{\DocFreqJoin{$kp_{mp}, O_1 \ldots O_n$}} {
        $O \gets \{\}$ \\
        \ForEach{$Enc(t_i) \in O_m$}{
             \ForEach{$Enc(t_j) \in O_n$} {
                \If{\IsMatch{$kp_{mp}, Enc(t_i), Enc(t_j)$}}{
                $O[t_j] \pluseq 1$;}
                \Else{$O[t_i] = 1$;}
             }
        }
        \KwRet $O$;
    }
    \BlankLine
    \KwIn{$kp_{mp}$: multiparty key-pair; $x, y$: (encrypted) token sets}
    \KwOut{$psi$: intersection size between $x, y$}
    \Fn{\PrivateSetIntersection{$kp_{mp}, x, y$}} {
    $psi \gets 0$ \\
    \ForEach{$Enc(t_i) \in x$}{
        \ForEach{$Enc(t_j) \in y$}{
            $psi \pluseq \IsMatch{$kp_{mp}, Enc(t_i), Enc(t_j)$}$ 
        }
    }
    \KwRet $psi$
    }
\end{algorithm}

\begin{algorithm}[tbh]
\caption{\texttt{he-ppjoin} ($R, t$) (with local \& global pre-processing)}
\label{alg:he-ppjoin}
    \KwIn{$kp_{mp}$: multiparty key-pair, $D_1 \ldots D_n$: local datasets; $O_1 \ldots O_n$: local document frequency mappings; $R$: (encrypted) record multi-set sorted by record length;  $O$: global (encrypted) document frequency mapping}
    \KwOut{$S$ with all pairs of records $ \langle x, y \rangle $, such that $sim(x, y) \geq t$}
    \tcp{Local pre-processing}
    \HiLi $kp_{mp} \gets \KeyGen{$kp_1 \ldots kp_n$}$\;
    \HiLi $D_1 \ldots D_n \gets$ \Encrypt{$kp_{mp}, t_i$ for $t_i \in r_j \in D_k$}\;
    \HiLi $O_1 \ldots O_n \gets$ \Encrypt{$kp_{mp}, t_i \in O_j$}\;
    \BlankLine
    \tcp{Global pre-processing}
    \HiLi $R \gets$ \{$D_1 \ldots D_n$\}\;
    \HiLi $O \gets \DocFreqJoin{$kp_{mp}, O_1 \ldots O_n$}$\;
    \BlankLine
    \tcp{PPJoin}
    $S \leftarrow \emptyset$\;
    $I_{w} \leftarrow \emptyset $\;
    \BlankLine
    \ForEach{$x \in R$}{
        $A \leftarrow$ empty map from record id to int\;
        $p \leftarrow |x| - \lceil t \cdot |x| \rceil + 1$\;
        \For{$i$ = 1 \KwTo $p$}{
            $w \leftarrow x[i]$\;
            \ForEach{$(y,j) \in I_{w}$ \textbf{such that} \\ \HiLi \IsMatch{$kp_{mp}, w, y[j])$ \textbf{and} $x,y \not\in D_k$ \\
            \textbf{and} $|y| \geq t \cdot |x|$}}{
                $\alpha \leftarrow \lceil \frac{t}{1+t}(|x|+|y|) \rceil$\;
                $ubound \leftarrow 1+min(|x|-i, |y|-j)$\;
                \uIf{$A[y] + ubound \geq \alpha$}{
                    $A[y] \leftarrow A[y] + 1$\;
                }
                \Else{
                    $A[y] \leftarrow 0$\;
                }
            }
            $I_{w} \leftarrow I_{w} \cup \{(x,i)\}$\;
        }
        \HiLi $\texttt{HE-Verify}(x, A, \alpha)$\;
    }
\end{algorithm}

\begin{algorithm}[t]
\caption{\texttt{HE-Verify} ($x, A, \alpha$)}
\label{alg:he-verify}
    \KwIn{$x, y$ are records containing encrypted tokens; $p_{x}$ is the prefix length of $x$ and $p_{y}$ is the prefix length of $y$; $O$ is the global (encrypted) document frequency mapping; $kp_{mp}$: multiparty key-pair}
    \ForEach{$y$ \textbf{such that} $A[y] > 0$}{
        \HiLi $Enc(x_l), Enc(y_l) \gets$ last tokens in resp. prefixes\;
        \HiLi \ForEach{$Enc(t) \in O.keys$} { 
            \HiLi \If{\IsMatch{$kp_{mp}, Enc(t), Enc(x_l)$}} { 
                \HiLi $w_{x}$ = pos. of $Enc(t)$ in $O$;
            }
            
            \HiLi \If{\IsMatch{$kp_{mp}, Enc(t), Enc(y_l)$}} { 
                \HiLi $w_{y}$ = pos. of $Enc(t)$ in $O$;
            }
        }
        $overlap \leftarrow A[y]$\;
        \uIf{$w_{x} < w_{y}$}{
            $ubound \leftarrow A[y] + |x| - p_{x}$\;
            \If{$ubound \geq \alpha$}{
                $overlap \leftarrow overlap$ +  \\
                \HiLi \PrivateSetIntersection{$kp_{mp}, x[(p_{x}+$ \\ \HiLi $1)\ldots|x|], y[(A[y]+1)\ldots|y|]$}\;
            }
        }
        \Else{
            $ubound \leftarrow A[y] + |y| - p_{y}$\;
            \If{$ubound \geq \alpha$}{
                $overlap \leftarrow overlap$ + \\
                \HiLi \PrivateSetIntersection{$kp_{mp}, x[(A[[y] +$ \\ \HiLi $1)\ldots|x|], y[(P_{y}+1)\ldots|y|]$}\;
            }
        }
        \If{$overlap \geq \alpha$}{
            $S \leftarrow S \cup (x,y)$\;
        }
    }
\end{algorithm}

\begin{figure*}[!t]
    \centering
    \includegraphics[width=0.9\linewidth]{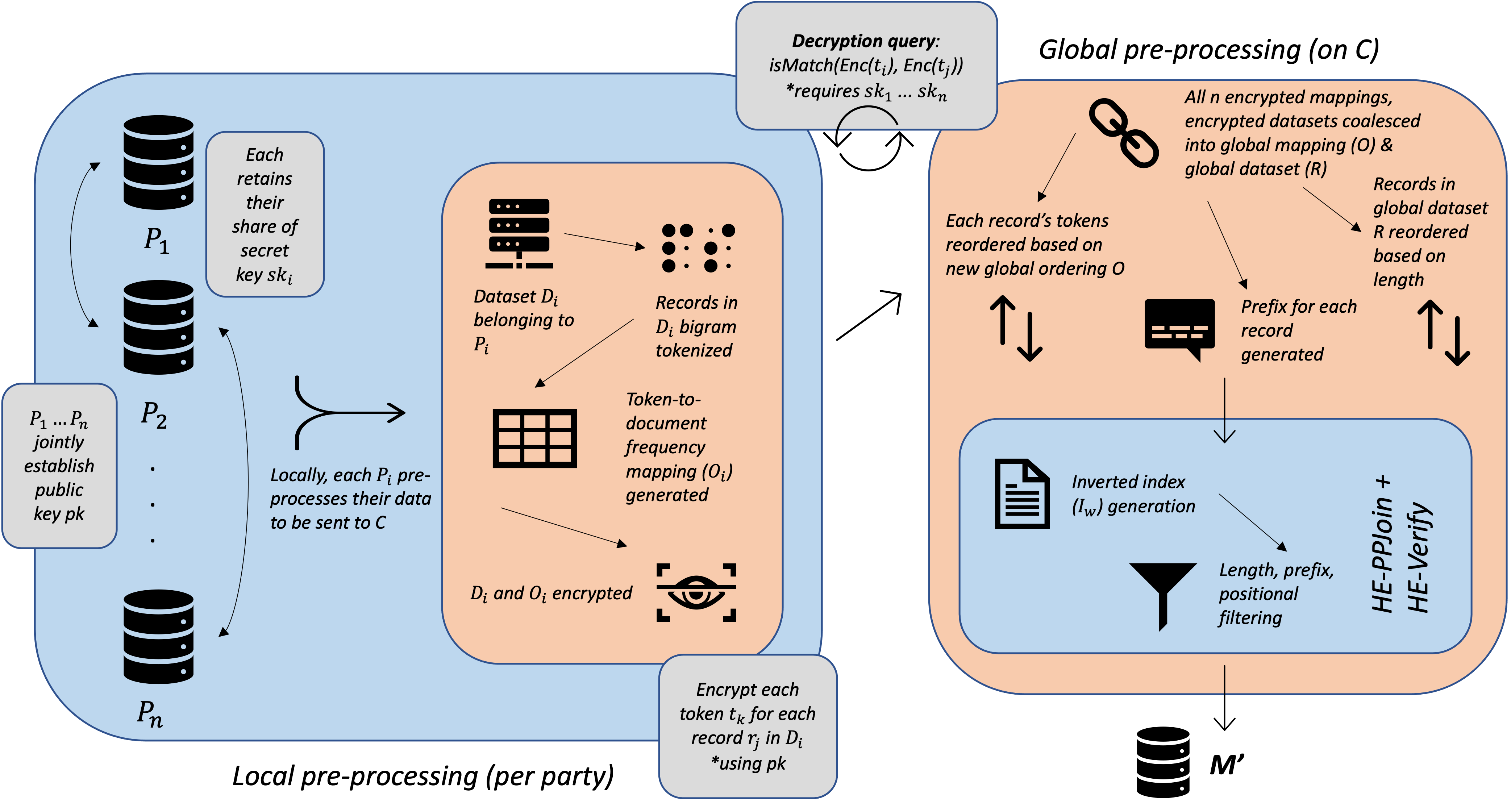}
    \caption{HE-PPJoin. Our protocol is split out into three main components: local pre-processing, global pre-processing, and the \texttt{he-ppjoin} \& \texttt{HE-Verify} algorithms. Local pre-processing is per-party utilizing threshold-HE with input $p_k$ (joint evaluation key), whereas the global pre-processing and the algorithms are performed over a semi-honest computation host $C$ utilizing $p_k$ and each party's encrypted data ($D_i, O_i$). \texttt{IsMatch} queries require a subset of all secret shares $s_1 \ldots s_n$.}
    \label{fig:he-ppjoin}
\end{figure*}

\subsection{Algorithm and protocol}
\label{sec:alg-protocol}

Our protocol (visualized in \cref{fig:he-ppjoin}) involves two forms of pre-processing before private versions of \texttt{ppjoin} and \texttt{Verify} (algorithms 1 and 2 in \cite{xiao2011efficient}) can be applied. \Cref{alg:he-ppjoin} and \cref{alg:he-verify} extend said algorithms to include local \& global pre-processing steps, data structure modifications, as well as other algorithmic adaptations. The explicit differences between our algorithms and their non-privacy preserving counterparts are highlighted in yellow for clarity. 

\noindent \textbf{Local pre-processing}: The first step for each party ($P_1 \ldots P_n$), is \textit{key generation}. Each party must first agree to the cryptographic scheme of choice (in this case BGV, in its threshold variant). Once consensus is established, parties interact to compute a joint public (evaluation) key that corresponds to the result of all their secret shares ($sk_1 \ldots sk_n$), each share retained by its corresponding party. This process is represented by line 1 in \cref{alg:he-ppjoin} via the \texttt{MultipartyKeyGen} (lines 1-3) procedure presented in \cref{alg:helper}; $kp_{mp}$ represents the multiparty key-pair containing the joint evaluation key $pk$ and the secret shares $sk_1 \ldots sk_n$. 

Once key generation is complete, each $P_i$ then proceeds to perform steps that mirror those of PPJoin's pre-processing: records in their dataset $D_i$ are \textit{bi}-gram tokenized and a token-to-document frequency map ($O_i$) is created. When processed, $D_i$ and $O_i$ are ready to be encrypted via $pk$: each token in each record in $D_i$ is encrypted, and likewise each key (token) in the map $O_i$ is encrypted (lines 2-3 in \cref{alg:he-ppjoin}). Each party then sends these locally encrypted data structures over to a computation host $C$. 

\noindent \textbf{Global pre-processing}: Once $C$ receives the $n$ pairs of encrypted data ($D_i, O_i$) from each party, they are coalesced into global data structures (lines 4-5 in \cref{alg:he-ppjoin}). Coalescing of all datasets into $R$ (the final sorted record multi-set) is quite simple -- all records from all datasets are simply clumped into one for $C$ to process. Coalescing all document frequency maps ($O_1 \ldots O_n$) into $O$, however, is a little trickier since all keys are now encrypted. In order to preserve token-content from $C$, we employ the \texttt{DocFreqJoin} helper method (lines 7-15 of \cref{alg:helper}) to securely tabulate the document frequency for all encrypted tokens amongst all maps. This procedure, in turn, utilizes another helper function \texttt{IsMatch} (lines 4-6 of \cref{alg:helper}) which can be seen as an interactive decryption query that performs a homomorphic subtraction operation between two tokens, multiplied by a uniformly random non-zero ciphertext element (generated on any of the data owners, and encrypted via $pk$) in order to determine if the plaintext tokens underlying have the same value. This query performs the distributed decryption ($DD$) algorithm that requires a subset of secret keys ($s_1 \ldots s_k$). Thus, for two tokens, $C$ receives back either 0 (if they match) or a uniformly random non-zero plaintext value (if they do not match), revealing no information about the encrypted tokens or their relationship with one another (in the case they do not match). Once $R$ and $O$ are both formed, each record's tokens are individually sorted based on the position assigned to them in $O$, prefixes for each record are generated, and records in $R$ are ordered by record size.

\noindent \textbf{Remark}: The \texttt{IsMatch} decryption query is the \textit{only} interactive procedure we allow $C$ to perform throughout the execution of the entire protocol. The use of the uniformly random ciphertext element randomizes the decryption such that $C$ cannot recover any relationship between $t_i, t_j$ unless they match (query returns 0), in which case the underlying value of the token is still obscured. Additionally, the use of \texttt{IsMatch} in ordered for-loops (e.g. in \texttt{DocFreqJoin}, \texttt{PrivateSetIntersection}) is parallelized for efficiency
as well as randomized for an undetermined pairing of tokens for comparison (such that $C$ can recover no relationship between tokens and the respective $O_i, O_j$ they came from). 

\noindent \textbf{\texttt{he-ppjoin} \& \texttt{HE-Verify}}: Finally, after pre-processing is complete, our privacy-preserving adaptations of \texttt{ppjoin} and \texttt{Verify} can be applied over the encrypted data seamlessly. In particular, line 14 of \cref{alg:he-ppjoin} applies \texttt{IsMatch} over a token from $x$'s prefix ($w$) compared with a token from record $y$ ($y[j]$). This operation is necessary to determine whether two encrypted tokens \textit{match} or overlap between the prefixes of two records in the inverted index $I_w$. Additionally, we apply an extra filtering step that does not exist in the original \texttt{ppjoin} algorithm: we require that $x$ and $y$ \textit{do not} belong to the same dataset $D_k$. This step is taken for the removal of extra comparisons and to remain consistent with our problem definition as it relates to ER -- we seek to find true pairs \textit{across} datasets (not those within the same dataset). While this formalism derives naturally from the concept of different data owners, this is not a concern in PPJoin due to its non-privacy preserving nature. 

In \cref{alg:he-verify}, the first main difference is in  determining $w_x, w_y$. When extracting out the last tokens in the prefixes of $x, y$, since they are encrypted, there is no way to directly get their document frequency or position from $O$. While all the tokens in $x, y$ naturally are a part of the key-set in $O$, it goes without saying that there is \textit{no} direct equality comparison possible between the two encrypted values, and so \texttt{IsMatch} is once again used to determine if their underlying values correspond to the same plaintext (lines 2-7). Besides this, the only other change necessary is in the calculation of the overlap between right-hand partitions of the $x, y$ (lines 13-14 and lines 19-20). To accomplish this, we apply \texttt{PrivateSetIntersection} (lines 16-21 of \cref{alg:helper}) to determine the number of overlapping elements between two inputted sets of encrypted tokens; the formulas for start and end indexes for the subsets of $x$ and $y$ match exactly to those of the original \texttt{Verify} algorithm.

\subsection{Correctness}
\label{sec:realworldcorrectness}

The correctness of our adaptation is built particularly on the correctness of BGV~\cite{cryptoeprint:2011:277} and threshold-HE~\cite{10.1007/3-540-44987-6_18, cryptoeprint:2017:257}. As mentioned in \cref{sec:he}, the use of threshold-HE replaces the algorithms $KeyGen$, $E$, and $D$ with their distributed, randomized counterparts: $DKG$, $DE$, and $DD$. Correctness can be \textit{guaranteed} if the result of applying the function $f$ homomorphically within the context of the distributed algorithms is indistinguishable from a non-private execution. 

We, first, utilize $DKG$ as part of local pre-processing (line 1 of \cref{alg:he-ppjoin}), where each party ($P_1 \ldots P_n$) interacts initially. It takes in the number of parties ($n$), and the threshold parameter ($d$). The corresponding output is a vector of secret keys ($s_1 \ldots s_n$) of dimension $n$, and a public evaluation key $pk$ (where all parties receive a key-pair ($pk, sk_i$)). $DE$ is also used during local pre-processing (lines 2-3 of \cref{alg:he-ppjoin}), where each party ($P_i$) encrypts their datasets ($D_i$) and their local document frequency map ($O_i$). The inputs are the party's secret key ($sk_i$) and a token ($t_j$), and the corresponding output is a ciphertext $c_j$. $DD$ is only used via the computation host $C$ during calls of the \texttt{IsMatch} (lines 14 of \cref{alg:he-ppjoin} and 4 \& 7 of \cref{alg:he-verify}) function. Being an interactive protocol that can be carried out by some subset of the total number of parties ($k < n$), the inputs are ($s_1 \ldots s_n$), the threshold parameter ($d$), and a ciphertext $c_j$. The corresponding output is a decrypted plaintext element $m_j$. However, in the context of our protocol, $DD$ is never used to directly decrypt the contents of any encrypted token; rather, we only allow decryption of the subtracted quantity of two ciphertexts multiplied by a randomized encrypted quantity for obfuscation (lines 4-6 of \cref{alg:helper}). This is required functionality needed to mimic the following filtering steps: where tokens are compared against each other (e.g. in the main loop of \texttt{ppjoin}) or to count overlapping elements between records (e.g. in \texttt{Verify}).

The \textit{correctness requirement}~\cite{cryptoeprint:2019:939}, in this case for threshold-HE, relies on the correctness of distributed decryption ($DD$): if at least $d$ parties do not perform adversarially, correctly following the denoted interactive nature of $DD$, then the resulting outputs are provably correct~\cite{10.1007/978-3-319-96884-1_19, cryptoeprint:2017:257}: $Pr[DD(sk_1 \ldots sk_n, d, f(DE(pk, m_1)) \ldots f(DE(pk, m_n))) = \mathcal{F}(m_1 \ldots m_n)] = 1$. That is, the result of applying $f$ (\texttt{he-ppjoin} \& \texttt{HE-Verify}) over encrypted tokens is \textit{indistinguishable} from the result of applying \texttt{ppjoin} \& \texttt{Verify} over plaintexts. 

\subsection{Privacy}
\label{sec:realworldprivacy}

In order to assess the resulting effects and leakage of privacy in a real-world execution of our protocol, we consider three distinct cases amongst the presence of a semi-honest (or \textit{honest-but-curious}~\cite{Hazay2010SemihonestA} adversary) $\mathcal{A}$ with potential collusion amongst entities: 
\begin{itemize}
    \item Case 1: $k < d \leq n$ data owners are compromised
    \item Case 2: only computation host is compromised
    \item Case 3: $k < d \leq n$ data owners \& computation host are compromised
\end{itemize}
where $k=$ the number of compromised entities, $d=$ the threshold parameter, and $n=$ the total number of protocol data owner participants. We assume $k$ to be \textit{strictly} less than $d$ in order to provide a bounded region in which we can guarantee privacy. This assumption is \textit{crucial} as it maintains the semantic security definitions presented in \cref{sec:he}: if $k \geq d$, $DD$ becomes a clear vulnerability as if the distributed decryption amongst data owners can no longer be trusted, in addition to the fact that $C$ has access to such queries (via \texttt{IsMatch}), correctness and privacy (especially given possible collusion amongst participants) becomes improbable to uphold. 

\noindent \textbf{Case 1}: Given $\mathcal{A}$ compromises $k < d \leq n$ data owners (while $C$ remains honest), semantic security is \textit{preserved}. This is due to the fact that even if the $k$ data owners collude, with only $k$ secret shares of $sk$, they cannot perform $DD$ together. Thus, as long as $k < d$, the execution of $DD$ operates as if all $n$ parties  were to be trusted, analogous to execution in a non-private scenario. Furthermore, since we assume a semi-honest setting, the $k$ compromised entities cannot deviate from the protocol in any manner (e.g. to perform \textit{active} attacks such as CCA). Given no power for active attacks or the power to manipulate distributed decryption maliciously, even with collusion, none of the $k$ data owners have an advantage in identifying any underlying token content. 
    
\noindent \textbf{Case 2}: Given $\mathcal{A}$ compromises the computation host $C$ (while all $n$ data owners remain honest), the only true power $C$ possesses is the ability to perform \texttt{IsMatch} queries (detailed in \cref{sec:alg-protocol}); the security of \texttt{IsMatch} relies on $DD$, meaning that without $d$ compromised, colluding data owners, the decryption will not leak any information. Additionally, a compromised $C$ must remain honest, and so no extra queries to the data owners can be made. In this way, $C$ is bounded by the information is receives from the data owners regarding token similarity (0 if they match, or an uniformly random value). Lastly, BGV also provides \textit{evaluation privacy}, meaning given a ciphertext $c_i$, the operations performed over it are additionally hidden from $C$. Because of this, $C$ cannot backtrack or reconstruct the corresponding plaintext element, giving it no additional advantage and \textit{preserving} semantic security. 
    
\noindent \textbf{Case 3}: Given $\mathcal{A}$ compromises $k < d \leq n$ data owners as well as the computation host $C$, the crux of the privacy-preservation argument comes as a result of what the $k$ data owners and $C$ can learn if they pool their views (collude) together. Once again, given that $k < d$, we know that there is no information leakage via $DD$ with respect to the \texttt{IsMatch} queries called from $C$. In this way, the views of $C$ (as described in case 2) as well as the compromised $k$ data owners (as described in case 1) are not any different as compared with their corresponding views if privacy was not a consideration. Since collectively the compromised $k$ data owners additionally have no adversarial power and the individual views offer no leakage of sensitive information, collusion between them is additionally futile.

While we claim "no leakage of sensitive information", it is important to formalize this by noting what we do allow our protocol to leak even given its privacy guarantees. We succeed in obfuscating record content as well as their corresponding tokenized representations, document frequency scores, prefixes, and sorted orderings. Additionally, just like in \texttt{ppjoin}, we do \textit{not} reveal any information regarding similarity scores of (non-)similar records. Furthermore, since $S$, the output, includes records that are \textit{similar}, it is not our goal to preserve knowledge of this from $C$. However, since \IsMatch is never fully applied directly over the entirety of two token sets (records) $r_i, r_j$ and only applied over encrypted prefixes, we do \textit{not} give $C$ knowledge regarding the specific similarity scores between them. We do, though, leak the size, or number of tokens in each record to $C$, since the use of encryption is applied at the granularity of tokens. Finally, we do \textit{not} consider the effects of possible inference attacks by $C$ ( tracing and performing access pattern reconstructions based on the discarding of non-matching values through the filtering procedures) because of it's inability to actively probe with arbitrary inputs and requirement to not deviate from the protocol given a semi-honest setting.

\section{Experiments}
\label{sec:experiments}

\subsection{Experimental environment \& settings}
\label{sec:experimentsetup}

Below, we provide an introduction into our environment settings, the platform in which we implemented our protocol, and datasets used along with evaluation metrics for quantifying our results. 

\noindent \textbf{Environment settings}: All our experiments were run on a virtual Ubuntu 18.04.4 LTS server, with 2 CPUs from Intel Xeon CPU E5-2690 v4 @ 2.60GHz and 4GB memory. We additionally take advantage of multi-threading provided via OpenMP~\cite{10.1109/99.660313}. 

\noindent \textbf{PALISADE}: We implemented our protocol in PALISADE~\cite{PALISADE} (version 1.11.6), an open-source, lattice based cryptographic library. The PALISADE library implements BGV in its respective RNS (\textit{residue number system}) variant \cite{heevaloftheaescircuit}, following the Homomorphic Encryption Standard~\cite{cryptoeprint:2019:939} by meeting proper bit-security thresholds, given via (semi-)automated parameters. Selecting appropriate values for these parameters is \textit{non}-trivial and is crucial to achieve the security and performance one would like to target. With respect to the BGV scheme, some parameters we configured: \textit{multiplicative depth} = 1, \textit{plaintext modulus} = 65537, \textit{sigma} = 3.2, \textit{security level} = 128 bits. With respect to threshold-HE, our experiments utilize an $n=2$.

\noindent \textbf{P4Join}: To the best of our knowledge, no existing implementation of P4Join is publicly available. Hence, we implemented P4Join~\cite{yixiang_yao_2020_3924703}
in Python, open sourcing it under MIT license. We use the following Python interpreters: CPython 3.8.10~\footnote{https://github.com/python/cpython}, the reference implementation of the Python interpreter, and PyPy 7.3.7~\footnote{https://www.pypy.org}, which is an optimized, compliant implementation using just-in-time (JIT) compilation.

\noindent \textbf{Datasets}: We utilize \textit{Febrl}, a synthetic dataset generated using \textit{dsgen}~\cite{christen2005probabilistic}, to evaluate run-time performance and to assess accuracy.
We choose Febrl due to it's synthetic nature, meaning that the dataset size and noise introduced is user-controllable. We prepare 2 different sized sets (100, 500) records (each in 3 different threshold values, $t$ = 0.2, 0.5, 0.8) for evaluation with the following splits for two data owners $D_1, D_2$: (20/80), (100/400).

\noindent \textbf{Evaluation metrics}: We define metrics commonly used to evaluate the performance and quality of an entity resolution algorithm. \textbf{Precision} ($pr$) = $\frac{M \cap P}{P}$ and \textbf{recall} ($re$) = $\frac{M \cap P}{M}$, where $P$ denotes the pair-sets found by the algorithm. In our experimentation, we explicitly use \textbf{F-score}, which combines these two metrics, defined as $\tfrac{2 \cdot pr \cdot re}{pr + re}$, operating as their \textit{harmonic mean}.

\subsection{P4Join performance verification}
\label{sec:p4join-experiments}
\begin{figure}[!t]
    \includegraphics[width=\linewidth]{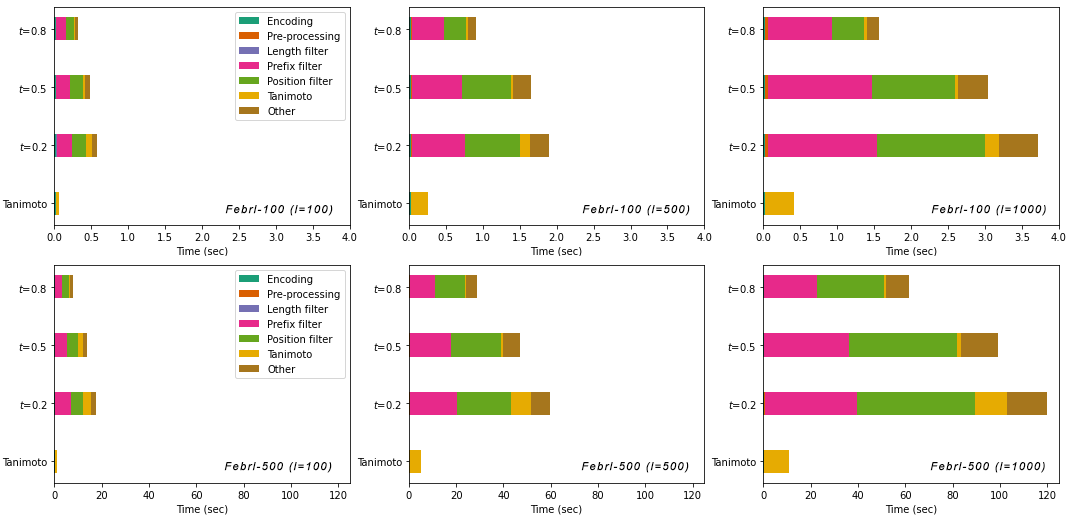}
    \caption{P4Join run-time on Febrl ($k=2$)}
    \label{fig:p4join-time-febrl}
\end{figure}

\noindent \textbf{Run-time} We profile the run-time cost of each P4Join component and compare them against pair-wise Tanimoto similarity as a baseline. In terms of P4Join parameters, we set the $k$ = 2 and $l$ = 100, 500 and 1000, and the results are shown in \cref{fig:p4join-time-febrl} for both Febrl-100/500. The cost of encoding, pre-processing and length filtering are trivial, whereas most of the run-time is spent on prefix filtering, position filtering and Tanimoto similarity. The three filters successfully eliminate more candidate pairs with a relatively large $t$, thus $t=0.8$ is visibly faster than $t=0.5$ and $t=0.2$. It is also noticeable that the longer signature (with greater $l$), the longer the run-time. While we set the maximum $l$ value to be 1000, this still does not guarantee the best accuracy. Tanimoto similarity, as part of P4Join, takes less time than a Tanimoto full comparison. Surprisingly, the pair-wise Tanimoto similarity runs drastically faster than P4Join. To explore this further, we additionally run such experiments with PyPy enabled (since \cite{sehili2015privacy} is in Java), as to exclude any added benefits received from from JIT. The overall run-time is improved, however, the relative relation between P4Join run-time and full Tanimoto run-time remains the same. Thus, in spite of the comparison savings achieved by the filters in P4Join, the overhead is \textit{significantly} larger than running Tanimoto directly.

\begin{figure}[!t]
    \includegraphics[width=\linewidth]{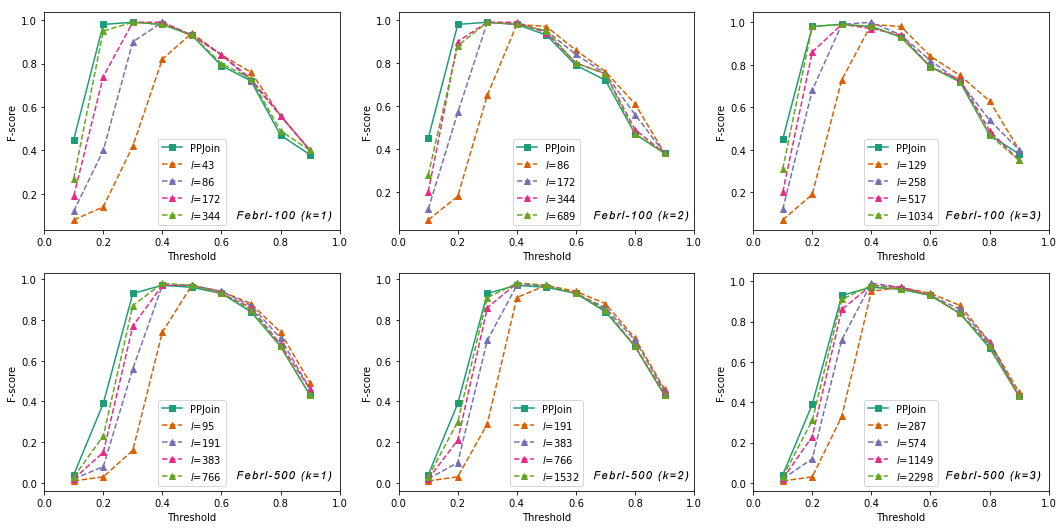}
    \caption{P4Join accuracy on Febrl}
    \label{fig:p4join-accuracy-febrl}
\end{figure}

\noindent \textbf{Accuracy} Choosing values for the parameters $k$ and $l$ involves a trade-off between accuracy and privacy. In this experiment, we count the total number of unique tokens and set $k$ to be 1, 2 and 3; the optimal $l$ is transformed from the equation for optimal $k$ along with $l/2$, $l/4$ and $l/8$. From \cref{fig:p4join-accuracy-febrl}, the results of P4Join have obvious differences as compared to PPJoin, especially when $l$ is far from the optimal value. Quite surprisingly, the F-scores are sometimes even better than in PPJoin. The reason behind this is that obfuscation via signatures can sometimes accidentally match records which are in fact not similar in PPJoin. Since the goal of P4Join is to simply adapt PPJoin to be privacy-preserving, this introduced uncertain-ness and arbitrariness as it relates to performance is unexpected. Another \textit{non-negligible} issue is the size of $l$. With Febrl-500, the optimal $l$ value  for $k=3$ is already $> 2000$. This implies that optimal usage (for accuracy) in the real world, requires extremely large signature lengths, which would inevitably slow down run-time performance. For example, assume \textit{bi}-gram tokenization on records formed by visible ASCII codes (from 32 to 126, 95 characters); the total number of unique tokens could thus be at most 9,025. If $k=3$, however, the optimal becomes $l=39,060$ which is not practical, and is even hard to quantify given some sacrifice with respect to accuracy.

\subsection{P4Join vs HE-PPJoin}

\begin{figure}[!t]
    \includegraphics[width=\linewidth]{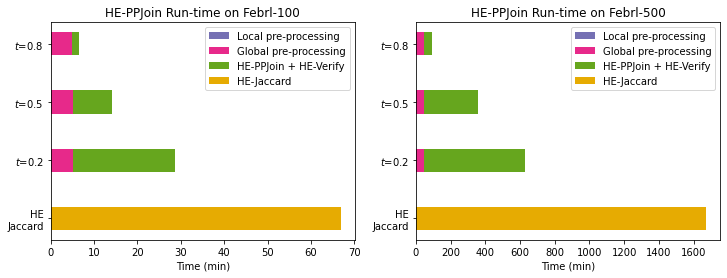}
    \caption{HE-PPJoin run-time}
    \label{fig:he-ppjoin-run-time}
\end{figure}



 We perform time-cost benchmarking for each of the three main components of HE-PPJoin on our two prepared Febrl datasets (Febrl-100, Febrl-500) utilizing three different threshold levels ($t=0.2, 0.5, 0.8$). The results are shown in \cref{fig:he-ppjoin-run-time}. We compare against a baseline full comparison performing HE-Jaccard, which refers to a HE-based private set intersection algorithm described in \cite{10.1145/3459637.3482318}.

For Febrl-100, we see that when $t=0.2$, the overall running time is around 30 minutes, whereas when $t=0.5$, the running time drops to approx. 15 minutes, and when $t=0.8$, it drops even further to less than 10 minutes. HE-Jaccard proves to be by far the most costly, taking over an hour to run all 1,600 ($ 20 \cdot 80$) comparisons. For Febrl-500, the same trend follows: when $t=0.2$, execution takes over 600 minutes, dropping to less than 400 minutes, and around 150 minutes for $t=0.5$ and $t=0.8$, respectively. HE-Jaccard for Febrl-500 again proves to be extremely costly, exceeding 1,600 minutes for all 40,000 ($100 \cdot 400$) comparisons. In terms of individual components, local pre-processing cost is negligible (in scale of approx. 30 seconds) as compared with the time cost of global pre-processing and the time for \texttt{he-ppjoin} and \texttt{HE-Verify}. Furthermore, the results show that overall, global pre-processing time remains constant regardless of the threshold value $t$, and purely depends on the dataset size as shown. The time cost for \texttt{he-ppjoin} \& \texttt{HE-Verify}, however, is strictly dependent on $t$ and can drastically improve (closer to $t=0.8$) or worsen (closer to $t=0.2$) given certain choices for it. 

Comparing with P4Join side-by-side, the added overhead of HE-PPJoin, incurred due to the private HE operators, is definitely significant. However, when comparing against baselines, it is important to note that P4Join does \textit{not} achieve expected run-time savings as compared with its own baseline (Tanimoto), whereas HE-PPJoin \textit{does}.

\noindent \textbf{Takeaway}: While the added cost of privacy preservation using HE is non-trivial, regardless of the choice for the threshold $t$, HE-PPJoin still shows \textit{significant} time cost savings as compared to its baseline full comparison approach (HE-Jaccard), meaning PPJoin's algorithmic benefits remain in-tact even in the private domain.

\subsubsection{Accuracy analysis}
\begin{figure}[!t]
    \includegraphics[width=\linewidth]{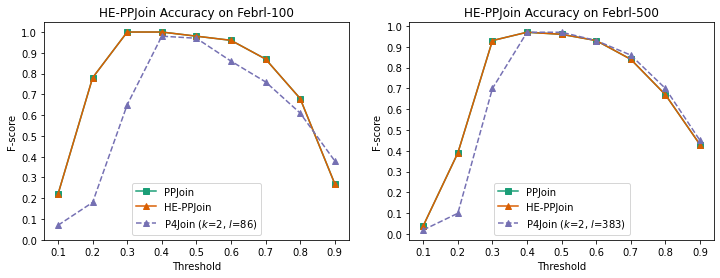}
    \caption{HE-PPJoin vs P4Join accuracy}
    \label{fig:he-ppjoin-accuracy}
\end{figure}

We additionally analyze the accuracy of our adaptation as compared with P4Join over both Febrl-100 and Febrl-500. As shown in \cref{fig:he-ppjoin-accuracy}, HE-PPJoin's F-score matches that of PPJoin's for all threshold $t$ ($0.1 \leq t \leq 0.9$) values. This is due to the fact that, unlike P4Join, we apply the usage of the three filtering techniques: length, prefix, and position in the exact same algorithmic format as that of the original PPJoin and Verify algorithms (as shown in \cref{alg:he-ppjoin}, \cref{alg:he-verify}). Furthermore, since BGV works with exact integer arithmetic, the results of $DD$ are exact matches to their corresponding plaintext values; this means that in checking for exact matches in \texttt{IsMatch}, there is no approximation error incurred. On the other hand, given P4Join parameter choices of $k=2$ and $l=86$, we can clearly see accuracy drop-off in Febrl-100 (for $0.1 \leq t \leq 0.4$ and $0.6 \leq t \leq 0.8)$ and in Febrl-500 (for $0.1 \leq t \leq 0.4$).

\noindent \textbf{Takeaway}: Our adaptation is able to not only maintain the performance benefits due to the algorithmic techniques present in PPJoin, but also does \textit{not} compromise the accuracy of the output (pairs returned as part of $M'$), no matter the threshold $t$ value.

\section{Conclusion and future work}
\label{sec:conclusion}

In this work, we present HE-PPJoin, a privacy-preserving adaptation of the PPJoin algorithm using the BGV fully-homomorphic construction, in it's threshold variant. We describe the run-time performance and accuracy of our approach, while also assessing the correctness and privacy of our adaptation in the presence of semi-honest adversaries in three different scenarios. Furthermore, we present a similar, rigorous analysis of these same properties as they relate to P4Join, an existing variant of PPJoin. Although our protocol introduces a non-trivial amount of overhead, we still see \textit{practical} performance execution for small, real-world dataset sizes, while ensuring \textit{no} information leakage. 

While the use of HE for work in applications such as PPER is still relatively new~\cite{10.1145/3459637.3482318} and inefficient~\cite{Armknecht2015AGT}, many recent works have established directions to bridge this gap for usability in the real world. For example,  \cite{cryptoeprint:2021:315, cryptoeprint:2021:1337} introduce 3x more efficient 
\textit{direct} comparison operators as well as faster \textit{special purpose} comparisons (e.g. less/greater than for sorting), whereas as \cite{cryptoeprint:2021:1412} establishes a new framework for \textit{non-interactive} multi-party HE setup. The use of such new directions in conjunction with our existing protocol can definitely be explored to scale to $n > 2$ party experiments, as well as complex, real-world datasets.

\bibliographystyle{unsrt}  
\bibliography{references}

\end{document}